# Fluid Guided CVD Growth for Large-scale Monolayer Two-dimensional Materials

Dong Zhou [a,b,#], Ji Lang [a,c,#], Nicholas Yoo [b,d], Raymond R. Unocic [e], Qianhong Wu [a,c,*], Bo Li [a,b,*]

[a] Department of Mechanical Engineering, Villanova University, Villanova, PA, 19085, United States

[b] Hybrid Nano-Architectures and Advanced Manufacturing Laboratory, Villanova University, Villanova, PA, 19085, United States

[c] Cellular Biomechanics and Sports Science Laboratory, Villanova University, Villanova, PA, 19085, United States

[d] Department of Chemical and Biological Engineering, Villanova University, Villanova, PA, 19085, United States

[e] Center for Nanophase Materials Sciences, Oak Ridge National Laboratory, Oak Ridge, Tennessee, 37831, United States

**ABSTRACT:** Atmospheric pressure chemical vapor deposition (APCVD) has been used extensively for synthesizing two-dimensional (2D) materials, due to its low cost and promise for high-quality monolayer crystal synthesis. However, the understanding of the reaction mechanism and the key parameters affecting the APCVD processes is still in its embryonic stage. Hence, the scalability of the APCVD method in achieving large scale continuous film remains very poor. Here, we use $MoSe_2$ as a model system and present a fluid guided growth strategy for understanding and controlling the growth of 2D materials. Through the integration of experiment and computational fluid dynamics (CFD) analysis in the full-reactor scale, we identified three key parameters: precursor mixing, fluid velocity and shear stress, which play a critical role in the APCVD process. By modifying the geometry of the growth setup, to enhance precursor mixing and decrease nearby velocity shear rate and adjusting flow direction, we have successfully obtained inch-scale monolayer $MoSe_2$. This unprecedented success of achieving scalable 2D materials through fluidic design lays the foundation for designing new CVD systems to achieve the scalable synthesis of nanomaterials.

# INTRODUCTION

Two-dimensional (2D) materials have tremendous potential in revolutionizing a wide range of industries, including but not limited to energy storage[1,2], health care[3,4], sensors[5], and actuators[6,7]. However, low-cost and scalable synthesis of high-quality 2D materials is a grand challenge for its commercialization. Among various synthetic methods for 2D materials, such as exfoliation [8-10], physical vapor deposition (PVD)[11-13], atomic layer deposition[14,15], metal-oxide chemical vapor deposition (MOCVD)[16], low-pressure chemical vapor deposition (LPCVD)[17-19], and atmospheric pressure chemical vapor deposition (APCVD)[20-22], APCVD shows the promise in providing high-quality and monolayer crystals using a simple and low-cost experimental setup. However, compared with the chemical process with self-limiting reaction mechanisms, e.g., growth of graphene on copper foil under low pressure[23,24], or those with a series of rigorously controlled sequences to deposit, remove and mix precursors, e.g., atomic layer deposition, the APCVD involves mixing multiple precursors under atmospheric pressure and thus cannot achieve controllable and scalable growth of nanomaterials. In the past decade, many studies have been dedicated to improving the control of the growth process, by adjusting the growth temperature[25-27], growth time[27], amount of precursor[28], precursor species[29,30], hydrogen content[31], substrate treatment[18,32], substrate species[33-35], and other factors[36]. Still, the multi-precursor APCVD process remains a black box where the reaction mechanism and knowledge of the controlling factors are not clear.

One of the key reasons behind the uncertainty of the APCVD process is that it involves chemical reactions coupled with complex fluid mixing where the transfer of momentum, heat, and mass significantly affects the reaction process and thereafter the final product[37]. Some preliminary theoretical and computational studies tried to correlate growth parameters with resultant 2D materials. For example, Xuan *et al.* developed a multiscale model for MOCVD at low pressure (200 Torr), where heat and mass transport equations at the reactor-scale were coupled with the mesoscale phase-field equations, to correlate the growth parameters with the distribution of synthesized $WSe_2$[38]. Fauzi *et al.* investigated the single precursor ($CH_4$) reaction model for the LPCVD process of graphene growth, where computational fluid dynamics (CFD) analysis was performed to obtain the temperature and velocity profiles during the deposition process[39]. However, these efforts, to the best of our knowledge, are mainly focused on explaining the experimental results. No active control mechanism has been achieved to tailor the growth of 2D materials, especially for the APCVD processes. In the paper, we, for the first time, develop a fluid guided growth (FGG) approach and use $MoSe_2$ as a demo system to understand and design the APCVD process. We have integrated chemical reaction with CFD analysis to simulate the mixing and reaction of precursors. The simulation, along with the experimental results, allows us to examine the role of the precursor mixing, the shear stress, and the fluid velocity in the outcome of the APCVD process, leading to the optimization of the growth crucible to achieve inch-scale monolayer film.

## RESULTS AND DISCUSSION

The schematic of the experimental setup and corresponding computer model for the $MoSe_2$ growth are presented in Figure 1a. Briefly, the samples are grown with solid $MoO_3$ and Se precursors using $Ar/H_2$ (85/15) as the carrier gas. The distance between the Se crucible and $MoO_3$ crucible is 7.5 in, and the growth temperature is set to be 760°C. In the growth process, the carrier gas enters the reaction tube from the inlet and carries the sublimated Se to react with $MoO_3$ and generate $MoSe_2$ crystals. The exhaust gas then leaves the tube via the outlet. The full reactor scale flow field, as well as the reaction process in the reaction quartz tube, is simulated by Ansys CFX®. A representative result of fluid velocity distribution in the quartz tube is shown in Figure 1b. We found that the structure and the position of crucibles can significantly change the fluid velocity distribution and hence influence the shear stress distribution, the concentration of the reactants and the corresponding chemical reaction, and thereafter affect the deposition process. Two traditional setups, namely standard APCVD (Figure 1c) and flipped APCVD (Figure 1d), were first used to elucidate the fundamental mechanisms governing the growth process. Correspondingly, two distinctive growth patterns have been observed experimentally, as shown in Figure 1f for standard APCVD and Figure 1g for flipped APCVD. The standard APCVD growth shows a vast empty region (lighter color) in the middle with little $MoSe_2$ deposition and a dense deposition (stripe purple color) close to the edge of the crucible. The flipped APCVD shows a more uniform deposition of $MoSe_2$ with some deposition along the centerline, but a more continuous deposition towards the edge of the substrate. The normalized cumulated $MoSe_2$ concentrations on the deposition substrates (the integration of the $MoSe_2$ concentration) solved by CFD are shown in Figure 1i for standard APCVD and Figure 1j for flipped APCVD. The color code of the figures is for the comparison of deposition magnitude for each of the figure itself. There is no correlation between the same color code for different figures. The simulation results are consistent with the experimental observation. Through the experimental and numerical study for standard and flipped APCVD, one finds that the mixing of the precursors, the fluid shear stress and velocity (including both magnitude and direction) near the substrate, are the essential factors for the synthesis of large-scale monolayer $MoSe_2$. This lesson was then applied to an evolved APCVD (Figure 1e) where we actively designed the APCVD setup, by creating controlled precursor mixing environment, decreasing shearing velocity, decreasing flow deviation to the substrate, and lowering shear stress over the growth substrate, to achieve the deposition of an inch-scale monolayer film (Figure 1h). The CFD simulation, demonstrating a uniform deposition over the substrate (Figure 1k), is consistent with the experimental observation. Overall, the CFD analysis can effectively reveal the mechanism of the growth process, and the modification of the flow field (or growth setup) can dramatically influence the deposition pattern.

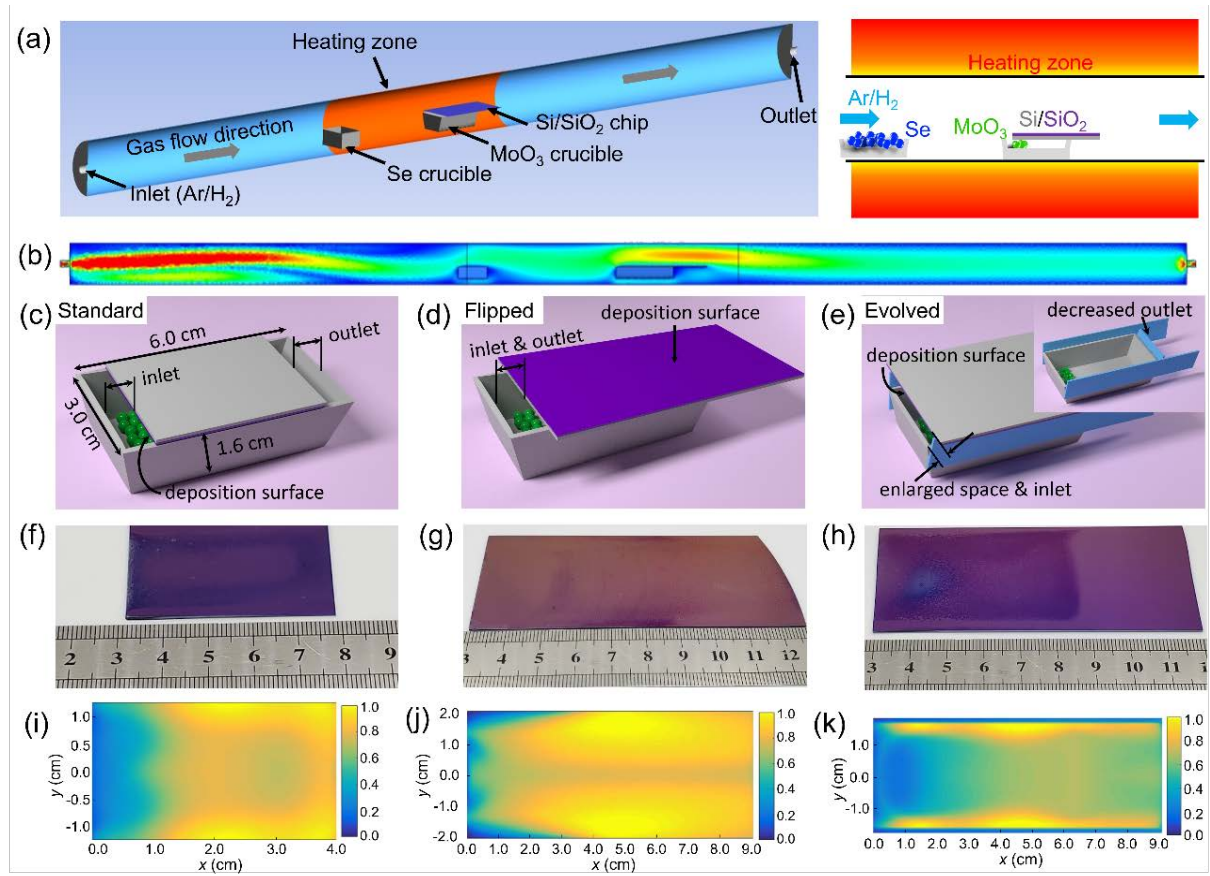

**Figure 1.** Fluid directed APCVD of MoSe$_2$ growth. (a) The structural model for CFD simulation in the full reactor length (left) and the schematic illustration of the experimental setup (right). (b) A representative figure is showing the velocity distribution in the reactor tube. (c)-(e) The growth setups for standard APCVD, flipped APCVD, and the evolved APCVD, respectively. (f)-(h) Real pictures of experimental samples using standard APCVD, flipped APCVD, and evolved APCVD, respectively. (i)-(k) CFD simulation of the cumulated concentration of MoSe$_2$ distribution, normalized with respect to the highest cumulated concentration over the corresponding substrate.

Figure 2 shows the experimental and numerical study for the standard APCVD process. For this setup, a silicon chip is placed on the MoO$_3$ crucible, with the polished thermal oxide (SiO$_2$) side facing the $MoO_3$ crucible. The silicon chip is 10 mm shorter than the length of the crucible, leaving two 5-mm openings at the two ends of the crucible for the carrier gas to flow into and out, Figure 2a. The Se enters the $MoO_3$ crucible and reacts with $MoO_3$ to form $MoSe_2$. Deposition, thereafter, appears on the bottom surface of the substrate, as shown in Figure 2a (right). The distribution of $MoSe_2$ on the substrate is observed using optical microscopy. Considering the symmetry of the growth setup, only half of the substrate is shown for both morphology observation and simulation. In Figure 2b, very few tiny triangular $MoSe_2$ crystals have been found along the central line ($y = 0$ cm). With $y$ increases, we can see the increase in the deposition concentration and crystal size. The more detailed morphology distribution can be found in Figure S1. We can also notice that the growth morphology changes from truncated triangles with straight edges to equilateral triangles with straight edges from the center to the edge. It suggests that the decreased concentration of Mo and increased concentration of Se from the center to the edge[40]. To make sure the triangular domains verified using optical images are indeed monolayer $MoSe_2$, we characterize them using TEM and Raman, photoluminescence (PL), and atomic force microscope (AFM), as shown in Figures S2-S4. From the TEM image and diffraction pattern (Figure S2a), we can clearly distinguish the monolayer character. Figure S2b shows a representative atomic resolution scanning transmission electron microscopy (STEM) image of the monolayer $MoSe_2$ synthesized by the standard APCVD. In Figure S3a, Raman vibration modes $A_{1g}$ can be seen at 241 cm$^{-1}$, which can be ascribed to the monolayer $MoSe_2$[41]. A prominent emission peak in the PL spectrum is located at 802 nm, which also confirms the direct bandgap of monolayer $MoSe_2$ at 1.55 eV (Figure S3b)[41]. The AFM morphology also confirms the monolayer of $MoSe_2$ with a thickness of 0.71 nm.

We numerically simulated the entire process that happens in the reaction tube. Figures 2c-2f present a snapshot ($t = 0.04$ s) of the simulation results during the reaction. The velocity distribution in the cross-section of the crucible, $y = 0$ cm, is shown in Figure 2c. The lower velocity under the substrate indicates that most of the carrier gas does not enter the crucible. It suggests that the sublimation of $MoO_3$ impeded Se from entering the crucible, and hence only a little amount of Se could join the reaction. Figure 2d presents the shear stress distribution over the deposition substrate, showing a lower shear stress region at the edge and a higher shear stress region at the center. Shear stress, the velocity gradient on the substrate that reflects the sweeping effect of the fluid flow, would drag the products to leave the substrate. Therefore, the lower shear stress distribution at the edge region favors the deposition process, which is consistent with the experimental result shown in Figure 2b. The $MoSe_2$ distribution over the substrate and in the cross-section of the crucible at $y = 0$ cm, are shown in Figures 2e and 2f, respectively. Before the sublimation of $MoO_3$, the $MoO_3$ crucible is filled with Se due to the continuous flow of the carrier gas. Once the sublimation of $MoO_3$ happens, the Se flow through the inner side of $MoO_3$ crucible is impeded, as demonstrated in Figure 2c for the velocity distribution, Figure S5 for the time-dependent propagation of Se, and Figure S6 for the time-dependent propagation of $MoO_3$. After the sublimation, the interrupted Se flow through the crucible is resumed, It enters the crucible, meeting and reacting with the remaining $MoO_3$. This explained why the concentration of $MoSe_2$ shows a non-uniform distribution with a narrow band in Figures 2e-f and Figure S7. The time-integration of the concentration field leads to the cumulated $MoSe_2$ concentration and is previously presented in Figure 1i. Intuitively, A higher cumulated concentration of $MoSe_2$ favors the growth locally, which is the reason why one observes a higher growth rate near the edge of the substrate, as shown in Figure 2b. Overall, the non-uniform distribution of the shear stress and the cumulated $MoSe_2$ concentration impede the growth of a continuous and uniform 2D film.

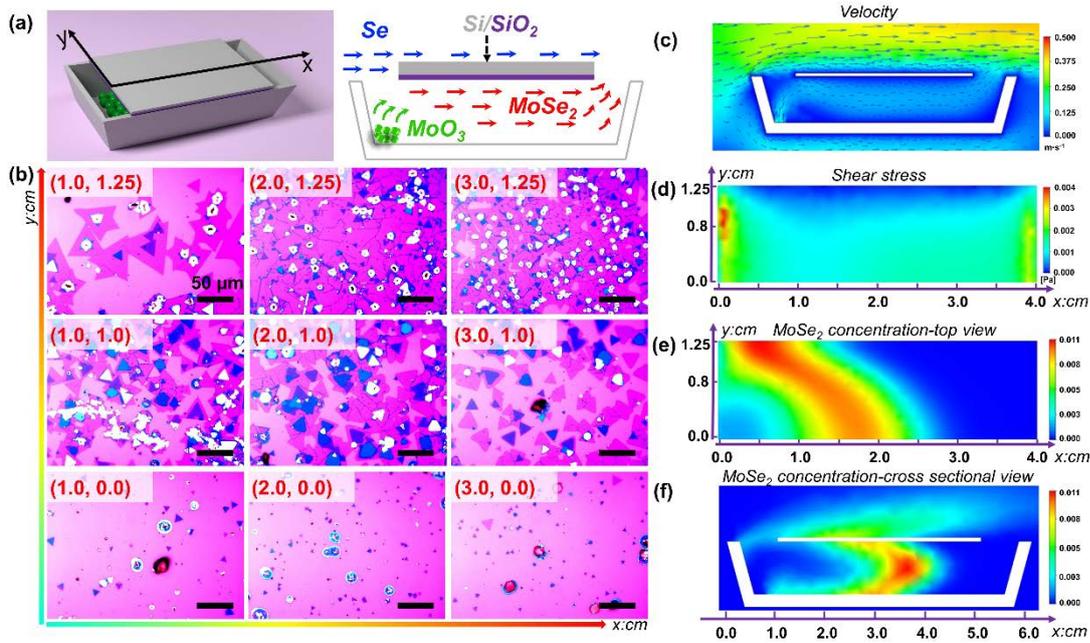

**Figure 2.** The standard APCVD growth. (a) Structural schematic of growth setup (left) and corresponding flow pathways for Se, $MoO_3$, and $MoSe_2$ (right). (b) Optical images of the morphology distribution over the growth substrate with 50 µm scale bar, the values shown in the images represent the corresponding coordinates over the substrate. (c) The velocity distribution and velocity vectors over the growth substrate at $t = 0.04$ s. (d) the shear stress distribution over the growth substrate at $t = 0.04$ s. (e) top view snapshot of $MoSe_2$ concentration distribution during the growth process at $t = 0.04$ s. (f) cross-section view of $MoSe_2$ concentration at $t = 0.04$ s. For (b)-(e), only half of the substrate is shown because of the symmetry of the growth setup.

Figure 3 shows the experimental and numerical study for the flipped APCVD process, where the silicon chip shown in Figure 2a is flipped upside down, with the polished thermal oxide ($SiO_2$) facing the outside of the $MoO_3$ crucible, as shown in Figure 3a. The silicon chip, 9 cm long and 4 cm wide, covered the right part of $MoO_3$ crucible, leaving a 5mm opening near the left edge as the outlet of the sublimated $MoO_3$. In this scenario, the inner space of the crucible serves as a Mo source. The Se, coming from the upstream, reacts with sublimated $MoO_3$ to form MoSe$_2$ outside of MoO$_3$ crucible. Deposition, thereafter, appears on the top surface of the substrate, as shown in Figure 3a (right). Similar to Figure 2, only half of the substrate is shown for both morphology observation and simulation. As shown in Figure 3b, along the central line ($y = 0$ cm), the products change from scattered particles ($x = 0$ cm to 2 cm), to small, sharp, straight-edge triangles ($x = 2$ cm to 6 cm), and to large irregulars ($x = 6$ cm to 9 cm) with hackly edges along the centerline, which reflects improved growth. With $y$ increases, significantly improved continuity of the deposited MoSe$_2$ can be found. The bare growth nearby the point at (9.0, 2.0) is caused by the degradation of a continuous MoSe$_2$ film, which is out of our discussion in this paper[42]. A more detailed distribution can be found in Figure S8. The energy-dispersive X-ray spectroscopy (EDX: Figure S9) mapping is used to determine the composition of the scattered particles. The mapping analysis shows that the atom ratio of Mo to Se is 10:1, which implies that little Se is involved in the reaction.

Figures 3c-3f present a snapshot ($t = 0.04$ s) of the simulation results during the reaction. The velocity distribution in the cross-section of the crucible, $y = 0$ cm, Figure 3c, indicates that the carrier gas encounters the jet of sublimated MoO$_3$. The mixture first flows away from the substrate due to the join of the MoO$_3$ and then gradually flows towards the substrate as the distance from the front edge increases. Therefore, it would be difficult for the products to attach to the substrate near the front edge of the substrate, which is the reason why there is almost no growth at locations (1.0, 0,0) and (1.0, 1.0) in Figure 3b, and why the growth at the center region improves along the flow direction, as shown in Figure 3b. Figure 3d presents the shear stress distribution over the deposition substrate. Two regions with lower shear stress are observed, one at the center near the front edge caused by the deviation of the fluid flow away from the substrate, and another near the rear edge and close to the tube wall where fluid velocity is low. As the flow deviation near the edge of the substrate was not severe as compared to the center, due to the constraint of the tube, the shear stress was relatively small, and the growth is better in the region, Figure 3b. The MoSe$_2$ distribution over the substrate and in the cross-section of the crucible at $y = 0$ cm, are shown in Figures 3e and 3f, respectively. In the flipped APCVD, the sublimated MoO$_3$ directly mixes with the Se, obtaining a better mixing, which is the reason why the MoSe$_2$ has a much larger area of higher concentration near the substrate (Figure 3e), as compared to the standard APCVD shown in Figures 2e and 2f. Figures S10, S11, and S12 show the time-dependent propagation of Se, MoO$_3$, and MoSe$_2$, respectively. The cumulated MoSe$_2$ concentration shown previously in Figure 1j, obtained from the time-integration of the concentration field, suggests a more uniform and slightly higher MoSe$_2$ concentration near the edge of the substrate. Hence, one observes a higher growth rate near the edge, consistent with the experimental result shown in Figure 3b. Overall, the distribution of the shear stress and the cumulated MoSe$_2$ concentration are much more uniform in the flipped APCVD, but, the flow deviation impeded the growth in the center region.

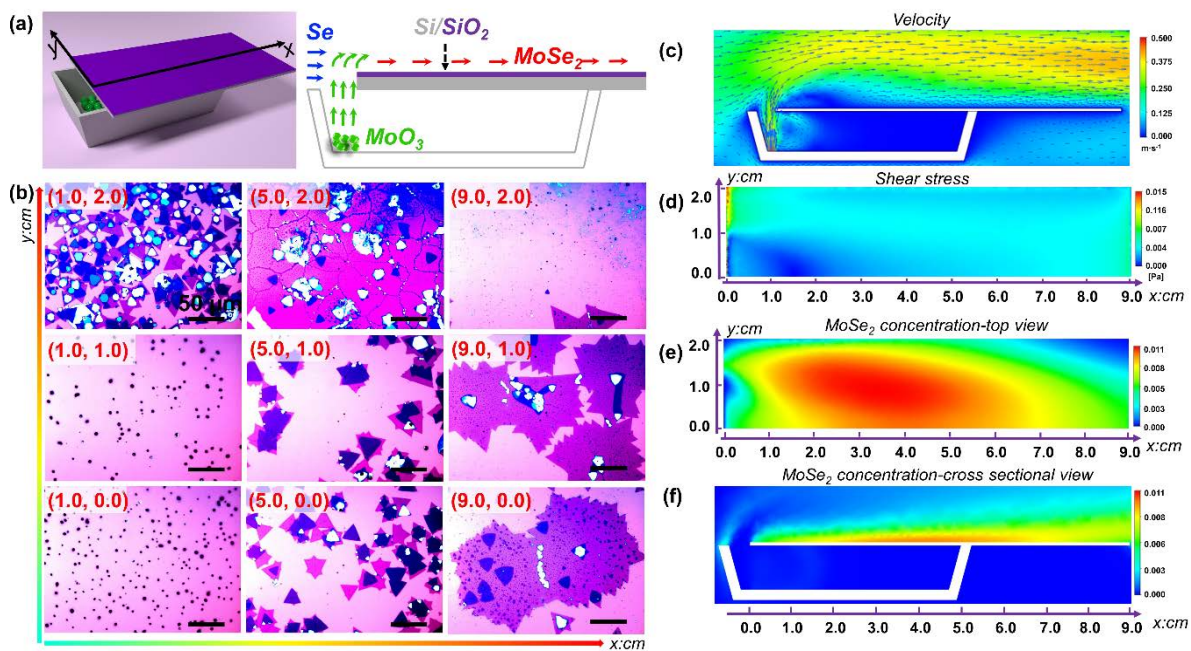

**Figure 3.** The flipped APCVD growth. (a) Structural schematic of growth set up (left) and corresponding flow pathways for Se, MoO$_3$, and MoSe$_2$ (right). (b) Optical images of the morphology distribution over the growth substrate with 50 µm scale bar, the values shown in the images represent the corresponding coordinates over the substrate. (c) The velocity distribution and velocity vectors over the growth substrate at $t = 0.04$ s. (d) the shear stress distribution over the growth substrate at $t = 0.04$ s. (e) top view snapshot of MoSe$_2$ concentration distribution during the growth process at $t = 0.04$ s. (f) cross-section view of MoSe$_2$ concentration at $t = 0.04$ s. For (b)-(e), only half of the substrate is shown because of the symmetry of the growth setup.

Lessons learned from the standard APCVD (Figure 2), and the flipped APCVD (Figure 3) suggest that, in order to achieve better growth, one needs to control the flow condition to achieve lower shear stress, better mixing, and preferred flow direction. This has inspired us to develop an evolved APCVD system, as shown in Figure 4. The deposition silicon chip was lifted by 1 cm from the crucible by attaching two side baffles (the blue colored components in Figures 1e and 4a) to have a larger inlet for the carrier gas to enter, and hence allowing direct mixing of Se and MoO$_3$. Another baffle is attached to the rear wall of the crucible (the blue colored component in Figures 1e and 4a) to create a narrow gap of 2 *mm* height as the gas outlet, which extends the precursors' dwell time under the substrate. The polished thermal oxide (SiO$_2$) side faces the MoO$_3$ crucible to avoid flow deviation from the substrate, as shown in Figure 4a (right). In Figure 4b, a continuous monolayer MoSe$_2$ film in the inch-scale is formed over the substrate in most regions except for a multilayer region near the point of (1.5, 0.0) and some scattered empty patches and add-on layers, *e.g.*, (3.5, 0.0). The monolayer quality also has been proved using the Raman and PL, as shown in Figure S13. Near the edge, one can clearly see decreased grain size along the flow direction, which changes from hundreds of μm to tens of μm. As $y$ increases, the continuity increases, reflecting an increased deposition. More detailed distributions can be seen in Figure S14.

Similarly, Figures 4c-4f present a snapshot ($t = 0.04$ s) of the simulation results during the reaction. The velocity distribution in the cross-section of the crucible, $y = 0$ cm, as shown in Figure 4c, suggests that carrier gas encounters the sublimated $MoO_3$ jet under the substrate. The mixture first flows towards the substrate, then becomes parallel to it, and subsequently deviates from the substrate when the gas flow approaches the exit between the substrate and the crucible. The exit acts as a throttle, near which the gas first flows towards the small gap, and then deviates from the substrate after it flows out of the crucible. At the center and close to the front edge, the straight flush to the substrate makes it easier for the products to attach to the substrate, consistent with the multilayer structure observed in Figure 4b at the location (1.5, 0.0). The deviation beyond the outlet was adverse for the deposition, which is the reason why there was almost no growth after the outlet, as shown in Figure 1h. The shear stress distribution over the deposition substrate, Figure 4d, indicates a lower shear stress region at the straight flush region, favoring growth in this region. The shear stress distribution is both low and uniform in all other regions except the one near the narrow outlet. The $MoSe_2$ distribution over the substrate and in the cross-section of the crucible at $y = 0$ cm, are shown in Figures 4e and 4f, respectively. For the evolved APCVD, Se enters the chamber between the crucible and the substrate, mixing and reacting with $MoO_3$. The mixing effect, including the high $MoSe_2$ concentration area, is between that of standard APCVD and flipped APCVD. Figures S15- S17 show the time-dependent propagation of Se, $MoO_3$, and $MoSe_2$, respectively. Overall, the well-controlled flow direction that is in parallel with the deposition substrate, the low and uniform shear stress in most of the region, and the reasonably good mixing of Se and $MoO_3$ in the evolved APCVD, lead to an improved cumulated $MoSe_2$ concentration as shown in Figure 1k, with a uniform concentration distribution in the center and a thin, higher cumulated concentration near the edge. Better growth is obtained as a result, as shown in Figure 4b. A continuous film is successfully obtained.

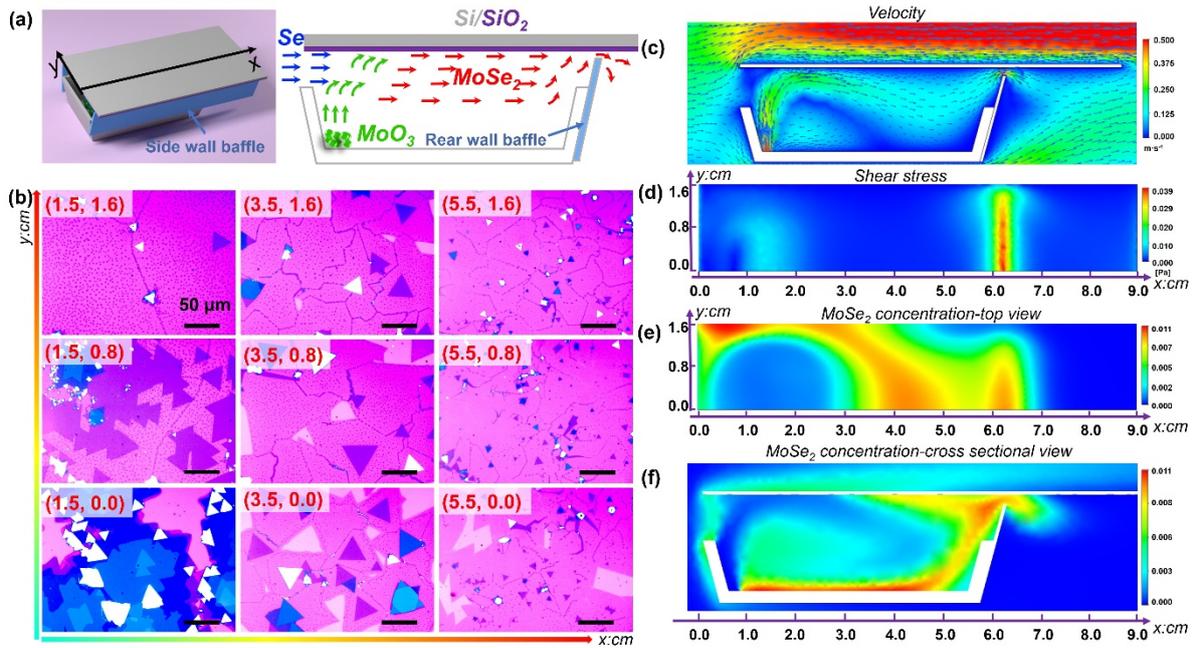

**Figure 4.** The evolved APCVD growth. (a) Structural schematic of growth set up (left) and corresponding flow pathways for Se, MoO$_3$, and MoSe$_2$ (right). (b) Optical images of the morphology distribution over the growth substrate with 50 µm scale bar, the values shown in the images represent the corresponding coordinates over the substrate. (c) The velocity distribution and velocity vectors over the growth substrate at $t = 0.04$ s. (d) the shear stress distribution over the growth substrate at $t = 0.04$ s. (e) top view snapshot of MoSe$_2$ concentration distribution during the growth process at $t = 0.04$ s. (f) cross-section view of MoSe$_2$ concentration at $t = 0.04$ s. For (b)-(e), only half of the substrate is shown because of the symmetry of the growth setup.

## CONCLUSIONS

In summary, we have achieved the growth of inch scale continuous monolayer MoSe$_2$ using an evolved APCVD by the design of the experimental setup to achieve optimal control of fluid flow. The CFD simulation analysis elucidates the mechanism involved in the APCVD process and guides the structure design and optimization. Our study shows that under the same experimental parameters, the change of the growth setup can significantly influence the evolution of the flow field and hence change the growth behavior. The factors that influence the APCVD process include but not limited to the precursor concentration distribution, fluid velocity (magnitude and direction), fluid shear stress over the substrate, as well as the mixing of Se and $MoO_3$. The higher and uniform concentration of precursors, low shear stress, and small flow deviation to the substrate can create a stable growth environment for large-scale monolayer 2D materials. We believe the improvement of APCVD to achieve high-quality growth should consider how to control the flow field. We anticipate that this principle is applicable for other 2D materials' growth, but to obtain controlled growth, the optimization should be made according to the properties of the specific precursors (*e.g.,* density, diffusivity, and reaction rate) and geometry of the growth setup.

## EXPERIMENTAL SECTION

**$MoSe_2$ growth.** The samples were grown by CVD with solid $MoO_3$ and Se precursors and using $Ar/H_2$ (85/15) as the carrier gas. All the reactions happened in a 2-inch diameter quartz tube under atmospheric pressure. The length of the quartz tube is about 47 inches. Growth substrates were Si with a 285 nm layer of $SiO_2$ and cleaned by using $N_2$. After being cleaned, one rectangular substrate was placed above the designed setup containing 15 mg of molybdenum oxide ($MoO_3$) powder (99.5%, Sigma-Aldrich). The $MoO_3$ crucibles for the three types of setups are the same, and the dimension and structure are shown in Figure S17. The exact location of the $MoO_3$ powder is directly close to the left end of the crucible. Another ceramic boat containing Se powder was placed at the upstream of the tube furnace at 300 °C. The distance between the Se boat and the $MoO_3$ boat was about 7.5 inches. The heating rate of all reactions was 50 °C·$min^{-1}$. After 10 min growth, the temperature was cooled down to room temperature naturally.

**CFD simulation.** To reveal the process that happened during the experiment, we use the commercial computational fluid dynamics (CFD) software package, Ansys-CFX®, to conduct the numerical simulation. The computational domain is shown in Figure 1a, which is as same as the real experimental setup. The time-dependent model is used to simulate the dynamic process of the CVD growth, and the total simulation time is 4 s. As the flow rate is low, the Reynolds number is under 100, and the flow is laminar. Besides, the growth process involves heat transfer. Therefore, the laminar flow and the total energy model are used in the simulation. The reaction among $MoO_3$, Se, and $H_2$ are simplified and simulated as a single-step, irreversible gas-phase reaction:

$$Se + 3H_2 + MoO_3 \rightarrow MoSe_2 + 3H_2O$$

In our simulation, we use $MoSe_2$ to represent the product produced in the gas phase and study how the concentration distribution, velocity, and shear stress can affect the deposition of the product. These principles also applicable if we use the intermediate phase $MoO_{3-x}Se_x$ (multi-step reaction)[42] to represent the product produced in the gas phase. Therefore, we simplify the whole reaction process to one equation and use $MoSe_2$ concentration for discussing. The reaction rate was assumed to be proportional to the minimal mol concentration of the reagent. The carrier gas, $Ar/H_2$ (85/15), is injected through the inlet with a constant flow rate of 0.0536 g·m$^{-3}$, which is the same as the experimental condition; $MoO_3$ and Se are injected from two crucibles downstream, respectively. In the experiment, Se, with a larger amount of 700 mg, began to sublimate at a lower temperature, and $MoO_3$, with a smaller amount of 15 mg, joined the reaction at a higher temperature. Therefore, in the simulation, it is assumed that, before the sublimation of $MoO_3$, the sublimation of Se has reached a steady state, which is the initial condition of the simulation. The sublimation of Se has a constant speed of 0.5 mg·s$^{-1}$ all through the simulation. On the other hand, the sublimation of $MoO_3$ has a constant speed of 15 mg·s$^{-1}$ only during the first 1s of the simulation, and $MoO_3$ stops sublimating in the remaining time.

These numbers are intended to provide physical insight and should not be considered as the exact number in the real case. The exhaust leaves the reaction tube through the outlet. For the remaining boundary conditions, the no-slip boundary condition is used for all the solid surfaces; all the walls in the heating zone are set to be 750 °C; all other walls are set to be 25 °C.

**Sample preparation and characterization.** The TEM sample was prepared using PMMA assisted method. Briefly, the PMMA solution was spin-coated on the surface of the as-grown monolayer $MoSe_2$ on a $Si/SiO_2$ substrate at 4000 rpm for 1 min. Then, the sample was baked at 120 °C for 5 min. After that, the $SiO_2$ layer was etched by the KOH solution (0.1 g/ml), and the detached $PMMA/MoSe_2$ film was transferred into the deionized water to remove residual KOH. The $PMMA/MoSe_2$ film was transferred to a TEM grid, and the PMMA was removed by the acetone and cleaned using isopropanol. The TEM image and selective area electron diffraction were obtained using a Hitachi H-7600 under the operation voltage of 100 kV. The annular dark-field STEM image was performed using an aberration-corrected Nion UltraSTEM100 operating at 60 kV. The morphology and distribution of as-grown $MoSe_2$ were characterized using optical microscopy (BX51M, Olympus CO., JAPAN). The EDX was conducted using the scanning electron microscopy (Hitachi S-4800). The thickness is characterized using atomic force microscopy (Agilent 5500). Raman spectroscopy and photoluminescence were conducted using a Renishaw instrument with an excitation wavelength of 532 nm.

# ASSOCIATED CONTENT

**Supporting Information**

The Supporting Information is available free of charge on the ACS Publications website.

# AUTHOR INFORMATION

**Corresponding Author**

qianhong.wu@villanova.edu (Q.W.)

bo.li@villanova.edu (B. L.).

**Author Contributions**

#D.Z. and J.L contributed equally to this work.

**ORCID**

Dong Zhou: 0000-0002-7387-2411

Ji Lang: 0000-0002-8886-1766

Nicholas Yoo: 0000-0002-3930-8606

Raymond R. Unocic: 0000-0002-1777-8228

Qianhong Wu: 0000-0002-6216-5674

Bo Li: 0000-0001-9766-7925

**Notes**

The authors declare no competing financial interests.

## Acknowledgments

The experimental study was supported by the startup fund of B. L.. Q. W. and J. L. were partially supported by the National Science Foundation CBET Fluid Dynamics Program under Award #1511096. D. Z., N.Y., and B. L. were partially supported by the U.S. Department of Energy, Office of Science, Office of Workforce Development for Teachers and Scientists (WDTS) under the Visiting Faculty Program (VFP). A portion of this research was conducted at the Center for Nanophase Materials Sciences, which is a DOE Office of Science User Facility.

Table of contents

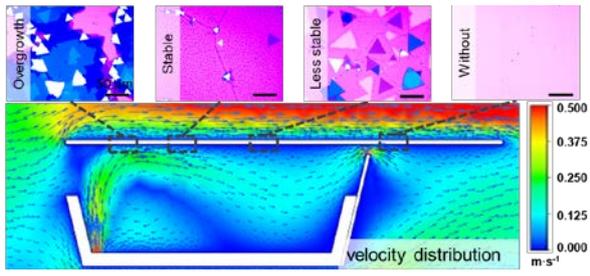